\documentclass[times]{elsarticle}
\usepackage[utf8]{inputenc}

\usepackage[margin = 1in]{geometry}
\usepackage{booktabs}
\usepackage{amsmath}
\usepackage{hyperref}
\usepackage{amssymb}
\hypersetup{colorlinks=true, linkcolor=blue, citecolor=blue, urlcolor=blue}
\usepackage[]{natbib}
\usepackage{soul}
\usepackage{multirow}
\usepackage{diagbox}
\usepackage{comment}
\usepackage{float}

\date{\today}

\begin{document}

\begin{frontmatter}
    \title{Photo-induced electronic excitations drive polymerization of carbon monoxide: A first-principles study}

    \author[label1]{ Rasool Ahmad \corref{cor1}}
    \ead{rasoolahmad1@llnl.gov}
    \author[label1]{Jonathan C. Crowhurst }
    \ead{crowhurst1@llnl.gov}
    \author[label1]{Stanimir A. Bonev}
    \ead{bonev@llnl.gov}
    \address[label1]{Lawrence Livermore National Laboratory, Livermore, California 94550, USA}
    \cortext[cor1]{corresponding author}

    \begin{abstract}
 Under pressure, carbon monoxide (CO) transforms into a polymer that can be recovered to ambient conditions. While this transformation can occur without additional stimuli, experimental observations have shown that laser irradiation can induce a similar transformation at reduced pressure. The resulting polymeric phase, which is metastable under ambient conditions, releases energy through decomposition into more stable configurations. Using time-dependent density functional theory and Born-Oppenheimer molecular dynamics simulations, we investigate the mechanism by which electronic excitation facilitates CO polymerization. Our calculations reveal that electronic excitation enhances carbon-carbon bonding, enabling polymerization at pressures significantly lower than those required by conventional compression methods. These findings suggest that a photo-assisted approach could be employed to synthesize novel, potentially energetic materials under less demanding pressure conditions.
    \end{abstract}        

\end{frontmatter}

\section{Introduction}

Carbon monoxide (CO) has attracted considerable attention from the scientific community as a potential energetic material. CO derives its energetic properties from the fact that molecular CO is metastable and kinetically transforms into a thermodynamically more stable polymeric phase under elevated pressures at room temperature. Notably, this high-energy-density amorphous polymeric phase can be recovered to ambient conditions. Its stored energy is released upon heating via the decomposition reaction: $2\text{CO} \longrightarrow \text{CO}_2 + \text{C}$. While theoretical calculations indicate that polymeric CO is not particularly energetic compared to other known materials~\citep{Bonev2021}, it remains a useful model system and potential precursor. Thus, methods for reducing the pressure required for its synthesis are worth exploring.

Since \citet{Katz1984} first reported evidence of polymeric CO, a growing body of experimental work\citep{Mills1985, Lipp1998, Lipp2005, Evans2006, Ceppatelli2009, Ryu2016} has focused on characterizing the structure and energy content of the resulting polymeric CO phase, as well as the thermodynamic conditions driving the molecular-to-polymeric transition process. Concurrently, several simulation and theoretical studies\citep{Bernard1998, Rademacher2014, Xia2017, Sun2011, Bonev2021} have aimed to understand the crystalline and polymeric CO structures through first-principles calculations. In particular, \citet{Bonev2021} performed first-principles molecular dynamics simulations designed to replicate typical experimental conditions, investigating the structure, thermodynamic conditions leading to the transition, and energy content of polymerized CO, and achieving results in good agreement with experimental data where available.

Moreover, several experiments~\citep{Lipp1998, Mills1985, Evans2006} have demonstrated that laser-induced photochemical conversion can significantly lower the pressure required for the molecular-to-polymeric transition. For instance, \citet{Evans2006} observed that intense laser irradiation at wavelengths of 488 nm (2.54 eV) and 514 nm (2.41 eV) reduces the transition pressure from 5.2~GPa to 3.2~GPa. Despite these findings, the underlying mechanism driving this photochemical transition and the resulting structures remains largely unexplored.

In this work, we investigate the effects of laser-induced electronic excitation on the molecular-to-polymeric transition of CO using first-principles molecular dynamics (FPMD), time-dependent density functional theory within the Tamm-Dancoff approximation (TDDFT/TDA), and linear response time-dependent density functional theory (LR-TDDFT) simulations. Our results convincingly demonstrate that electronic excitations enhance bonding between the carbon atoms of CO molecules, thereby inducing the molecular-to-polymeric transition at lower pressures. Additionally, we observe that compression shifts the absorption spectrum of molecular CO towards lower energies, suggesting that high-energy visible or ultraviolet laser light may be sufficient to excite electrons and facilitate the transition.

The manuscript is organized as follows: Section~\ref{sec:comp_details} provides a detailed description of the computational methodologies used in this study, including FPMD, TDDFT/TDA, and LR-TDDFT simulations. Section~\ref{sec:results} presents the simulation results. Finally, Section~\ref{sec:conclusion} summarizes the key findings, discusses them in the context of existing experimental results, and identifies potential directions for future research to further investigate and utilize the phenomena described in this work.

\section{Computational details}
\label{sec:comp_details}
This section outlines the computational methods employed in this study. First-principles molecular dynamics (FPMD) simulations were performed using the \texttt{VASP} software. Time-dependent density functional theory (TDDFT) combined with the Tamm-Dancoff approximation (TDA) was utilized as implemented in the open-source code \texttt{ORCA}. Additionally, linear response time-dependent density functional theory (LR-TDDFT) simulations were conducted using the open-source code \texttt{SALMON}.

\subsection{FPMD simulation}
FPMD simulations were carried out using \texttt{VASP}\citep{Kresse1993} to investigate the isothermal compression of CO at a temperature of 300~K. The system consists of 64 CO molecules within periodic cubic simulation cells. All FPMD calculations were performed using the Perdew-Burke-Ernzerhof generalized gradient approximation (PBE-GGA)\citep{Perdew1996}, employing a single $\Gamma$- centered $\boldsymbol{k}$-point sampling of the Brillouin zones. Four- and six-electron projector augmented wave (PAW) pseudopotentials~\citep{Blochl1994}, with core radii of 1.50 and 1.52 Bohr for C and O, respectively, were utilized, along with a plane-wave cutoff of 550~eV. Finite-temperature FPMD simulations were conducted within the canonical $NVT$ensemble (constant number of particles, $N$, volume, $V$, and temperature, $T$), where $T$ was controlled by a Langevin thermostat. The timestep for ionic motion was set to 1 femtosecond.

The simulation cell, shownin Figure\ref{fig:structure_CO}, was initially equilibrated in the molecular phase at a temperature of 300~K and a density of 1.12~g/cm$^3$. Subsequently, the system was progressively compressed by reducing the volume of the simulation cell by 1\% increments, followed by equilibration of the compressed cell for a duration of 0.1 picoseconds after each step.

We now describe the procedure for employing FPMD simulations to investigate the qualitative effects of photo-induced electronic excitation on the molecular-to-polymeric transition. In a typical Born-Oppenheimer FPMD simulation timestep, the electronic wavefunction is first calculated by solving Schr\"odinger's equation, treating the ions as fixed at their instantaneous positions. Next, only the lowest electronic bands (valence bands) are populated with the appropriate number of electrons. Finally, the energy and ionic forces are determined, followed by the integration of the equations of motion for the ions. Throughout the simulation, the electrons remain in the ground state corresponding to the instantaneous ionic positions. This conventional approach to FPMD is hereafter referred to as ground-state molecular dynamics (GSMD) simulation.

To mimic the effects of excited electrons, after obtaining the ground state occupancy of electronic bands in a well-equilibrated system,  electrons are partially removed from a few of the highest occupied bands (valence bands) and transferred to the lowest unoccupied bands (conduction bands). Following this protocol, three excited electron states are constructed by progressively transferring more electrons from valence to conduction bands. These states are referred to as Level 1, Level 2 and Level 3 excitations. At the initial density of 1.12 gm/cm$^3$, the  excess energies  associated with Level 1, Level 2 and Level 3 excitations are  2.9 eV, 2.8 eV and 8.0 eV, respectively, relative to the GSMD structure. These excess energies can be achieved through the application of visible laser light. The three  levels of  excitation enable us to explore the trends arising from the effects of electronic excitation.

\subsection{TDDFT/TDA calculation}
TDDFT/TDA calculations\citep{Hirata1999} were performed to explicitly investigate the influence of excited electrons on the intermolecular bonding between two CO molecules, thereby rationalizing the results obtained from FPMD simulations with excited electrons. For these calculations, we considered two CO molecules in a vacuum. The molecules were arranged such that the two C atoms faced each other while the two O atoms faced away from each other (see Figure~\ref{fig:tddft_result}). The distance between the two C atoms was gradually varied, and the positions of the two O atoms were relaxed while keeping the C atoms fixed. For each configuration of the CO molecules—defined by the separation between the two C atoms—the energies of the ground state and the next nine excited states were determined. All TDDFT/TDA calculations were conducted with \texttt{ORCA}~\citep{Neese2020}, using the PBE-GGA exchange-correlation functional and the def2-tzvpp basis set for all electrons in the system.

\subsection{LR-TDDFT calculation}

LR-TDDFT calculations were employed to determine the optical absorption spectrum of the molecular CO system~\citep{Sander2017}. In this method, we first computed the ground-state electron density and the Kohn-Sham orbitals, which serve as the starting point for the subsequent TDDFT calculations. At time $t=0$, a weak spatially uniform external impulsive electric field is applied, and the Kohn-Sham orbitals are evolved by solving the time-dependent Kohn-Sham equations. This evolution generates a time-dependent current density in the system, whose Fourier transform yields the frequency-dependent dielectric function. The imaginary part of the dielectric function is the optical absorption coefficient of the material under investigation.

To elucidate the effect of compression on the optical absorption spectrum, LR-TDDFT computations were performed on 64 CO molecules at two densities: low (1.12~g/cm$^3$) and high (1.7~ g/cm$^3$). Both systems were in the molecular phase. LR-TDDFT simulations were carried out using the open-source real-space TDDFT code \texttt{SALMON}~\citep{Noda2019}. The calculations utilzed a $80\times\ 80\times 80$ grid in real space and employed the PBE-GGA exchange-correlation functional within the adiabatic approximation. The Kohn-Sham equations were integrated with a timestep of 10$^{-3}$ femtosecond for a total simulation duration of 20 femtoseconds.

\section{Results}
\label{sec:results}
This section describes the results obtained from the FPMD, TDDFT/TDA and LR-TDDFT calculations.

\subsection{FPMD calculations}
\label{sec:fpmd_result}

\begin{figure}[ht!]
    \centering
    \includegraphics[width=\linewidth]{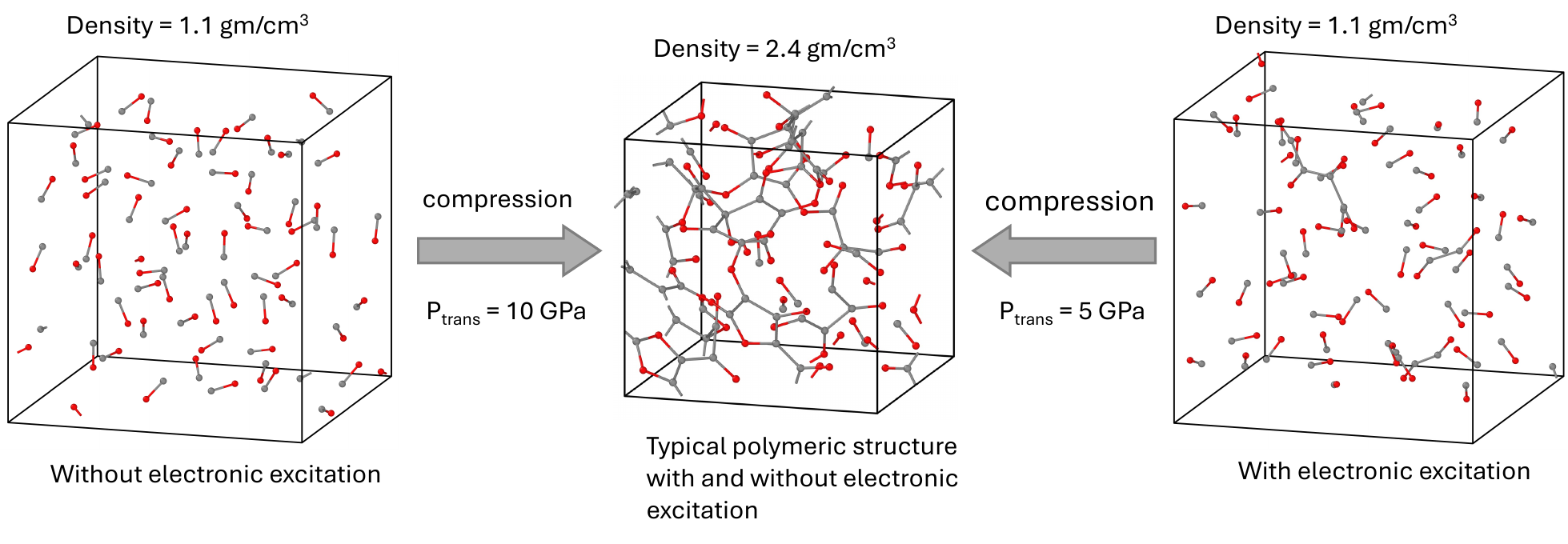}
    \caption{Atomic structures of 64 CO molecules at initial density of 1.12~g/cm$^3$ and after isothermal compression to a density of 2.4~g/cm$^3$ at 300 K. The structure shown on the left is at the initial density and without any electronic excitation, which is in the pure molecular phase. On the right is the structure at the initial density but with electronic excitation, which contains bonds between carbon atoms. In the middle is a typical polymeric structure obtained after isothermal compression up to 2.4~g/cm$^3$, for both cases with and without electronic excitation. The transition pressures (P$_\text{trans}$) are also indicated for both cases. Gray atoms represent carbon and red atoms denote oxygen.}
    \label{fig:structure_CO}
\end{figure}

\begin{figure}[ht!]
    \centering
    \includegraphics[width=\linewidth]{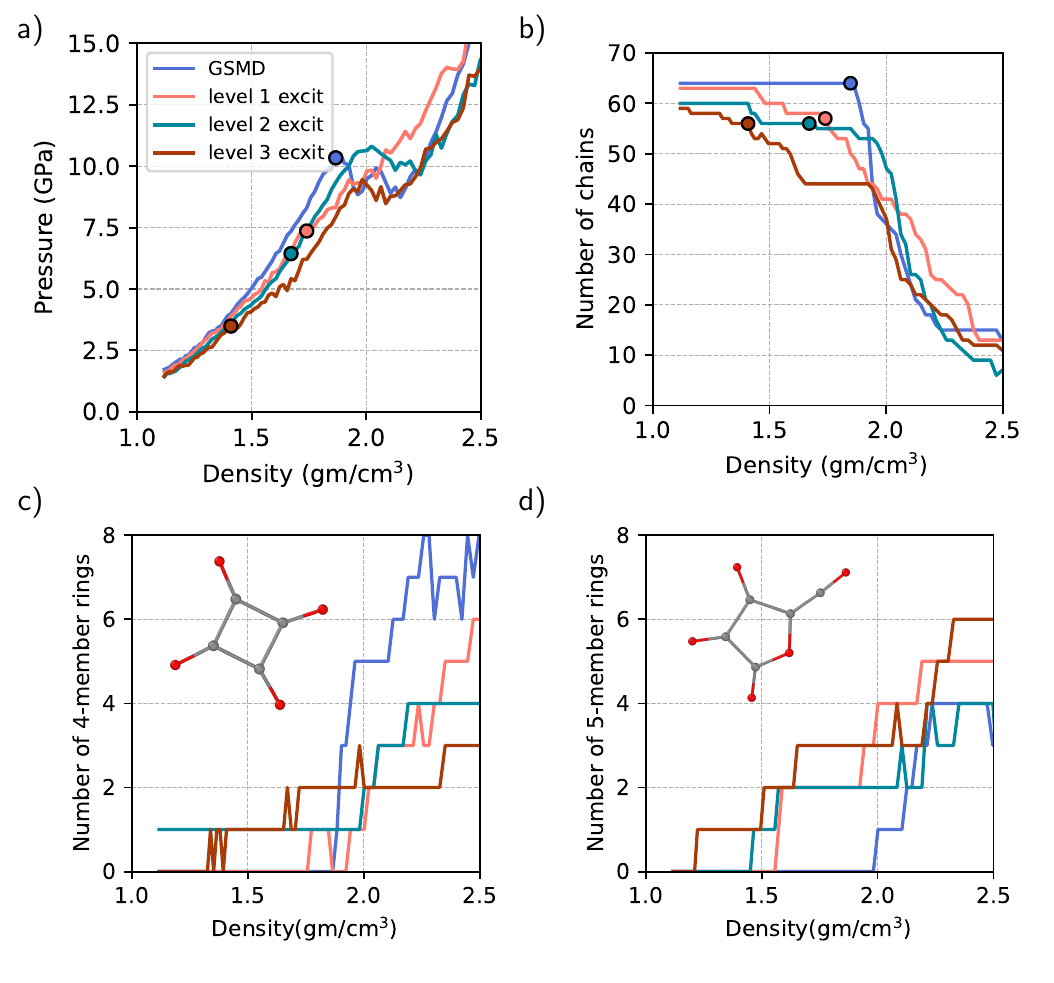}
    \caption{Results from FPMD simulations of isothermal compression of CO molecules at 300~K. The simulations were performed under four conditions: a standard Born-Oppenheimer FPMD simulation where electrons occupy the ground state corresponding to ionic positions (GSMD), and three levels of electronic excitation as described in the text. (a) Evolution of pressure as a function of density. (b) Change in the number of chains/molecules versus density. Circular markers on the curves in (a) and (b) indicate the approximate onset of the molecular-to-polymeric transition. (c) Number of 4-member rings versus density. (d) Number of 5-member rings versus density. Each plot includes four curves corresponding to the simulation conditions: blue for GSMD, orange for level 1, green for level 2, and red for level 3 electronic excitation. Panels (c) and (d) also illustrate typical 4-member and 5-member rings formed during compression. Gray atoms represent carbon and red atoms denote oxygen.}
    \label{fig:CO_figure}
\end{figure}

We first present the results of 300~K isothermal compression of 64 CO molecules obtained from FPMD simulations performed under four levels of electronic excitation. Figure~\ref{fig:CO_figure}(a) shows the variation of pressure with density as the systems undergo polymerization. The four curves are color-coded based on the levels of electronic excitation: blue for no excitation (GSMD), orange for Level 1 excitation, green for Level 2 excitation, and red for Level 3 excitation. Initially, the system begins in the molecular phase at a pressure of 1~GPa and a density of 1.12~g/cm$^3$. As compression increases, both density and pressure rise across all four excitation levels. However, upon further compression, a noticeable drop in pressure occurs, signaling the onset of the molecular-to-polymeric transition. 

This pressure drop is most pronounced in the GSMD case (blue curve), where the transition is relatively sharp and occurs at a density of approximately 1.8 gm/cm$^3$ and a pressure of $\approx$10~GPa. This sharp transition is further highlighted in Figure~\ref{fig:CO_figure}(b), which shows the evolution of the number of separate molecules or chains as a function of density. For GSMD, there is a steep decline in the number of chains at a density of 1.8 gm/cm$^3$, clearly marking the molecular-to-polymeric transition. 

The polymeric phase formed under compression is not entirely linear and includes ring structures, as shown in Figure\ref{fig:structure_CO}. Figures\ref{fig:CO_figure}(c) and (d) show the number of 4-member and 5-member rings that emerge during compression. For GSMD, the appearance of rings begins at a density of 1.8 gm/cm$^3$, coinciding with the onset of polymerization. 

\begin{figure}[ht!]
    \centering
    \includegraphics[width=\linewidth]{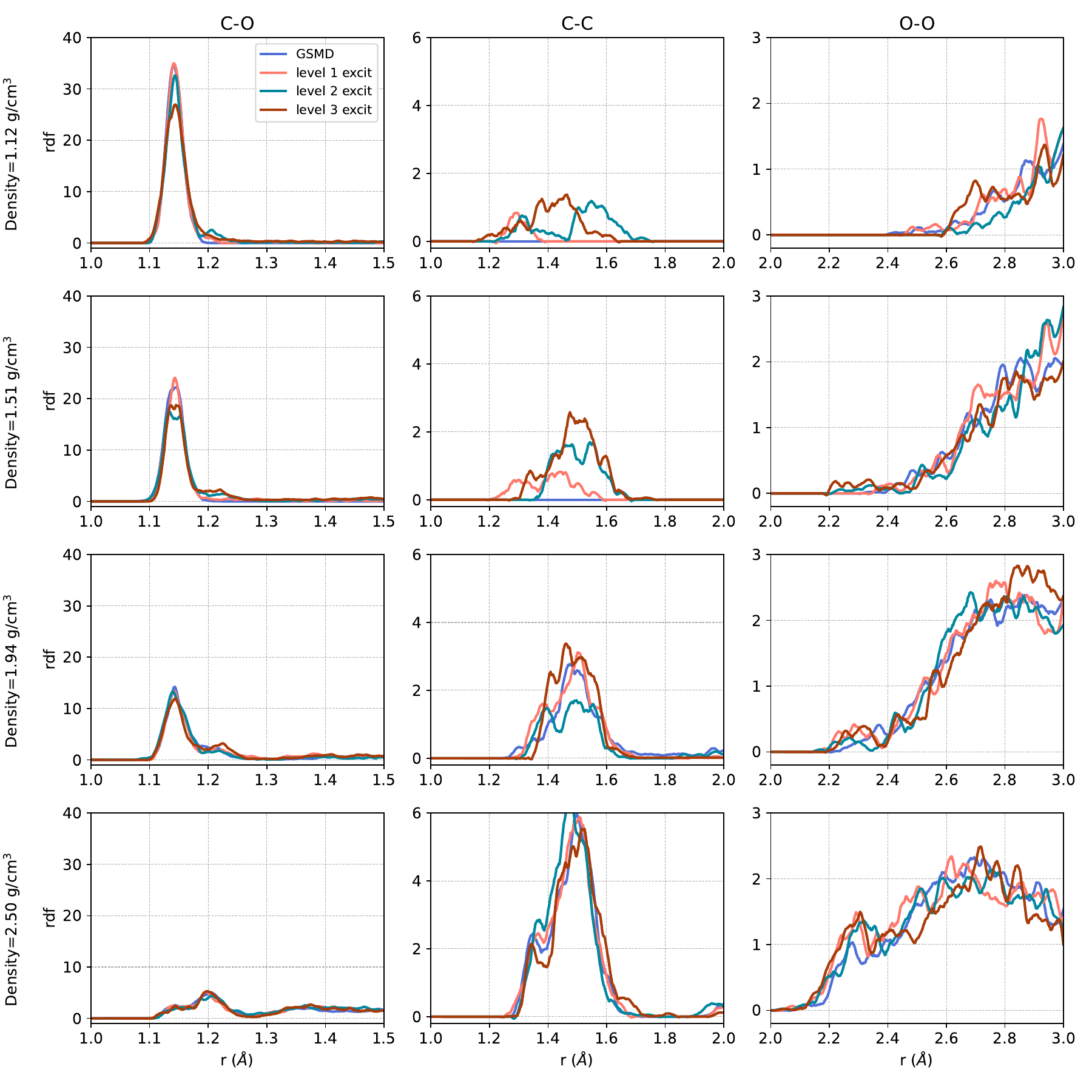}
    \caption{Radial distribution functions of C-O, C-C, and O-O atom pairs at various densities during the isothermal compression of CO molecules at 300~K. Each plot  includes four colored curves representing the different simulation conditions considered in this work: blue for GSMD, orange for level 1 electronic excitation, green for level 2 electronic excitation, and red for level 3 electronic excitation.}
    \label{fig:rdf_CO}
  \end{figure}

Furthermore, we show in Figure~\ref{fig:rdf_CO} the evolution of the partial radial distribution function (rdf) of C-C, C-O and O-O pairs as a function of density for both GSMD and the electronic excitation cases. At the initial density of 1.12~gm/cm$^3$, the GSMD case contains only CO molecules, as evidenced by the sharp peak in the C-O rdf at $\approx$1.12~\AA and the flat lines in the C-C and O-O rdf plots up to 2.5~\AA. At a density of 1.94 gm/cm$^3$, the GSMD case exhibits a peak in the C-C rdf plot at $\approx$1.5~\AA, indicating the formation of C-C bonds and the transition into the polymeric phase. As density increases further, the C-O rdf peak progressively diminishes while the C-C rdf peak strengthens, reflecting the extent of the molecular-to-polymeric transition.

With increasing levels of electronic excitation, the molecular-to-polymeric transition shifts towards lower densities and pressures. The transition with electronic excitation, however, exhibits a smoother character compared to GSMD, spanning a wide range of densities. Figure~\ref{fig:CO_figure}(b) shows that the starting configurations with electronic excitation, equilibrated at 300~K temperature and a density of 1.12~gm/cm$^3$ contain fewer than 64 chains (the starting number of CO molecules), indicating  the combination of two or more CO molecules even at low densities and pressures (see Figure~\ref{fig:structure_CO}).

The smooth nature of the transition is is further evident from Figure~\ref{fig:CO_figure}(a), where the pressure drops observed in the electronic excitation cases
are much less pronounced compared to the sharp drop seen in the GSMD case near the transition density. Additionally,  Figures~\ref{fig:CO_figure} (a) and (b), reveal that ring structures in the electronic excitation cases begin to form at significantly lower densities and pressures than in the GSMD case.

Furthermore, in all three levels of electronic excitation, as depicted in Figure\ref{fig:rdf_CO}, C-C bonds of length $\approx$1.5\AA~ are noticeable even at the initial density of 1.12~g/cm$^3$. The number of C-C bonds, as expected, increases with compression as polymerization proceeds. At a density of  1.5~g/cm$^3$ and a pressure of $\approx$ 5~GPa, the system contains both polymeric and molecular phases in significant amounts. Notably, the structures recovered by decompressing the system preserves the polymeric phase with rings formed during the compression.

\subsection{TDDFT calculations}
\label{sec:tddft_result}
The results from the FPMD simulations presented in Section~\ref{sec:fpmd_result} demonstrate that electronic excitations enhance bonding between carbon atoms, thereby assisting the transition from the molecular to the polymeric phase. However, in these simulations, electronic excitation was modeled by artificially transferring electrons from low-energy to high-energy electronic bands. The fraction of electrons placed into high-energy bands was not determined based on a rigorous approach.

\begin{figure}[ht!]
    \centering
    \includegraphics[width=\linewidth]{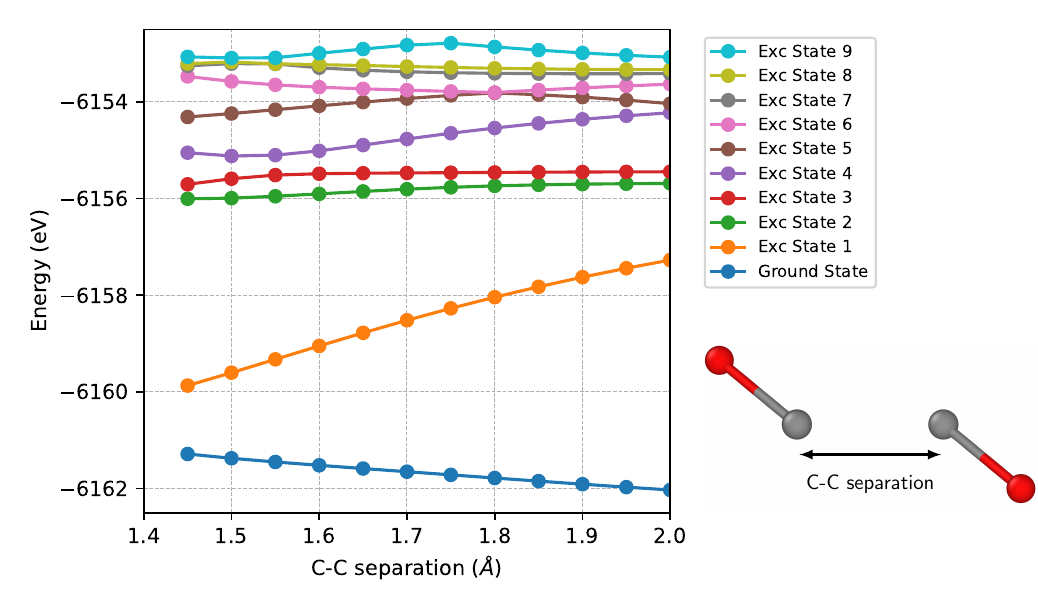}
    \caption{Ground- and excited-state energies of two CO molecules as a function of the distance between the two carbon atoms, obtained from TDDFT/TDA calculations. A schematic configuration of the two CO molecules is also shown: the two carbon atoms face each other, while the two oxygen atoms lie outside, facing away from each other. Gray atoms represent carbon, and red atoms represent oxygen. The TDDFT/TDA calculations did not converge for separations less than 1.45~\AA  between the two carbon atoms.}
    \label{fig:tddft_result}
\end{figure}

We now present the results obtained from TDDFT/TDA calculations for a system of two CO molecules. As illustrated in Figure\ref{fig:tddft_result}, the two CO molecules are positioned such that their carbon atoms face each other, while their oxygen atoms point outward, away from one another. The energies of the ground state and the next nine excited states were calculated as functions of the distance between the two carbon atoms, and the results are plotted in Figure\ref{fig:tddft_result}. The separation between the two carbon atoms ranges from 1.45 to 2.0~\AA, with increments of 0.5~\AA. It is worth noting that the TDDFT/TDA calculations did not converge for separations smaller than 1.45 \AA.

In the ground state, the combined energy of the two CO molecules unsurprisingly ncreases as the separation between the molecules decreases (blue curve in Figure~\ref{fig:tddft_result}), highlighting the intrinsic stability of the CO molecule. Interestingly, however,  the energy of the first excited state declines as the two CO molecules are gradually brought closer together (orange curve in Figure~\ref{fig:tddft_result}). Although we are not able to obtain well-converged TDDFT/TDA results for C-C separation smaller than 1.45 \AA, it is reasonable to assume that, if the molecular distance continues to decrease, the energy of the first excited state would eventually begin to rise again due to the dominating effect of interionic repulsion at very short distances.  Consequently, in the first excited state, the two molecules would achieve a bounded stable state. Evidence of such a stable state can be observed in the fourth excited state (violet curve in Figure\ref{fig:tddft_result}) at approximately 1.5~\AA, and in the sixth excited state (pink curve in Figure\ref{fig:tddft_result}) at around 1.8~\AA. The results presented in this section demonstrate that electronic excitations promote bonding between carbon atoms and, thus, facilitate the molecular-to-polymeric transition of CO-based materials.

\subsection{Optical absorption coefficient}
\label{sec:opt_abs_results}

% visible light range: 1.63 to 3.1 electron volts (eV)
% uv light energy range: between 3.1 and 12 electron volts (eV)

The spectrum of the laser energy (photon energy) that can be absorbed by the molecular phase of CO is determined by employing linear-response TDDFT as implemented in \texttt{SALMON}. Two structures $-$ with densities of 1.12~g/cm$^3$ and 1.70~g/cm$^3$ $-$ are chosen prior to the onset of  polymerization, specifically to demonstrate the effect of compression on modifying the spectrum of the absorbed laser energy.

\begin{figure}[ht!]
    \centering
    \includegraphics[width=\linewidth]{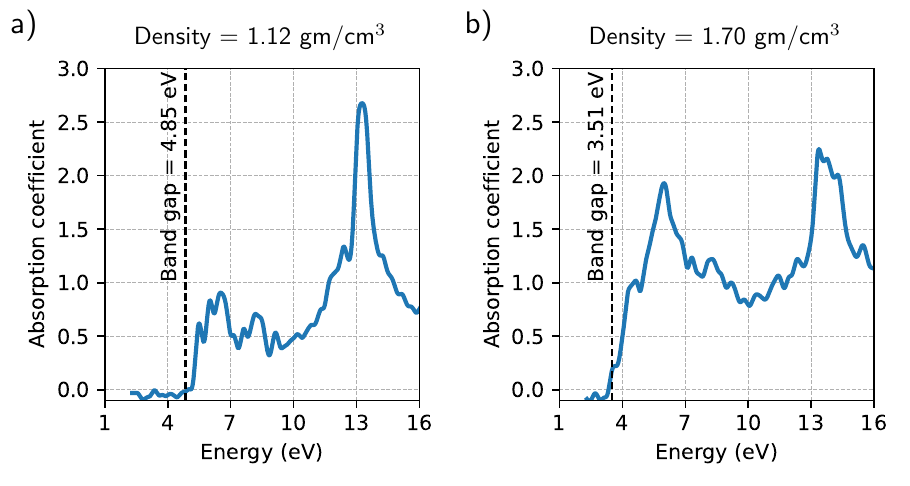}
    \caption{
Optical absorption coefficients for the molecular phase of CO at two different densities: (a) 1.12~g/cm$^3$ and (b) 1.70~g/cm$^3$, calculated using linear response Time-Dependent Density Functional Theory (TDDFT). Vertical lines indicate the electronic band gap in each plot.}
    \label{fig:opt_abs}
\end{figure}

Figures~\ref{fig:opt_abs} (a) and (b) show the optical absorption spectrum for densities of 1.12~g/cm$^3$ and 1.70~g/cm$^3$, respectively. The plots also include lines corresponding to the electronic band gaps at the respective densities. At the lower density of 1.12 gm/cm$^3$, the structure absorbs photons with energy greater than 5.0~eV, reflecting the electronic band gap of 4.85~eV. At the higher density of 1.70~g/cm$^3$, the lowest photon energy absorbed by the structure is around 4.0~eV, corresponding to the electronic band gap of 3.51 eV. Both optical absorption spectra exhibit a pronounced peak at around~13 eV. Additionally, the optical absorption spectrum of the higher-density structure (Figure\ref{fig:opt_abs}(b)) contains an additional peak near 6.0~eV, which is weakly present in the lower-density structure (Figure~\ref{fig:opt_abs}(a)).

Thus, the results presented in this section show that electron in CO-based molecular systems can be excited by the application of ultraviolet laser light. However, with compression, the electronic  band gap, decreases, enabling the molecular CO system to absorb lower-energy photons. This suggests the possibility for utilizing visible light (photon energy range from 1.6 to 3.1~eV) to excite electrons and accelerate the polymerization process.

\section{Discussion and conclusions}
\label{sec:conclusion}

In this work, we investigate the effects of electronic excitations on the molecular-to-polymeric transition of CO by performing a series of FPMD, TDDFT/TDA and LR-TDDFT simulations.  The primary finding of thisstudy, as demonstrated by FPMD simulations, is that electronic excitations can induce bonding between carbon atoms, thereby facilitating the onset of molecular-to-polymeric transition at lower pressures than those required under ground-state conditions. TDDFT/TDA calculations further support the FPMD results by showing that excited states promote the formation of bonds between carbon atoms in CO molecules. Additionally, LR-TDDFT calculations reveal that compression reduces the laser energy threshold for optical absorption in the CO system, suggesting that pressure enhances the efficiency of laser-induced electronic excitations. 

The experimental study by \citet{Evans2006} reported that photochemical reactions induced by a 2.54~eV laser lead to the polymerization of CO at 3.2~GPa, significantly lower than the 5.2~GPa required in the absence of laser irradiation. Raman spectroscopy further revealed that the structure recovered from higher pressures (5.2~GPa) without laser exposure lacks any detectable signatures of CO molecules, whereas the structure recovered from lower pressures (3.2~GPa) under laser irradiation contains a notable amount of CO molecules alongside the polymeric phase. These experimental observations align well with the two key qualitative conclusions of this work, namely: (1) the polymerization pressure decreases with increasing level of electronic excitation, and (2) the molecular-to-polymeric transition occurs smoothly in the presence of laser irradiation, with both molecular and polymeric phases coexisting across a wide range of densities and pressures, whereas the laser-free molecular-to-polymeric transition is more abrupt.

The quantitative results presented in this work show some deviation from experimental observations.For instance, the molecular-to-polymeric transition of the CO system without electronic excitations occurs at approximately 10~GPa in FPMD simulations, which is slightly higher than the experimentally observed range of 5–7~GPa~\citep{Evans2006, Lipp2005}. This difference can be attributed to the use of the PBE-GGA exchange-correlation functional, which is known to overestimate transition pressures. Importantly, \citet{Bonev2021} demonstrate that more accurate transition pressures can be obtained using hybrid exchange functionals such as Heyd-Scuseria-Ernzerhof (HSE)\citep{Heyd2003}. Nevertheless, the ensembles generated using PBE-GGA and HSE functionals are found to be highly consistent with one another, validating the use of the computationally efficient PBE-GGA functional in this study to capture the qualitative trends in the molecular-to-polymeric transition.

The accuracy of the computed optical absorption spectrum depends on the choice of the exchange correlation functional~\citep{Sander2017}. Application of computationally more expensive hybrid functionals could provide more accurate results compared to the PBE-GGA functional. Nonetheless, we expect that the observed trend of the absorption spectrum shifting towards lower photon energies with compression would remain consistent even with the application of more accurate hybrid functionals.

In the FPMD simulations with excited electrons conducted in this work, electrons were manually transferred from valence to conduction bands. While this approach is effective for gaining qualitative insights into the effects of electron excitations, it does not systematically incorporate the influence of laser parameters, such as intensity and energy. A possible direction for future research would be to explicitly model the interaction of laser and the coupled dynamics of ions and electrons, which can be achieved within the framework of Ehrenfest dynamics~\citep{Ojanpera2012}. Although several TDDFT codes have implemented Ehrenfest dynamics, their applications have primarily focused on small molecular systems, and further testing is required to ensure their reliability for larger bulk systems.

% These findings may have broader and more general applicability in molecular-to-polymeric transition of other energetic materials. For example, molecular N$_2$ transforms into high-energy-density polymeric phase at extreme conditions, and a suitably chosen laser can potentially be employed to reduce the transition pressure. 

In conclusion, the results presented in this work provide compelling evidence that electronic excitations play a critical role in promoting molecular-to-polymeric transitions. These findings may have broader applicability to similar transitions in other energetic materials~\cite{Mattson2004, Ciezak-Jenkins2017, Pravica2017}, and suggest a promising pathway for synthesizing polymeric CO-based materials under milder conditions.

\section{Acknowledgments}
This work was performed under the auspices of the U.S. Department of Energy by Lawrence Livermore National Laboratory under contract number DE-AC52-07NA27344. Authors acknowledge funding support from the Laboratory Directed Research and Development Program at LLNL under the project tracking code 23-ERD-028.

\addcontentsline{toc}{section}{References}
\bibliographystyle{model1-num-names}

\end{document}